\documentclass[showpacs,prb,floatfix,superscriptaddress,twocolumn,amsmath]{revtex4-1}
\usepackage[latin1]{inputenc}
\usepackage[english]{babel}
\usepackage{epsfig}
\usepackage{amssymb}  
\usepackage{amsmath}  
\usepackage{latexsym} 
\usepackage{amsthm}   

\usepackage{graphicx}       
\usepackage{epstopdf}
\usepackage{dcolumn}        
\usepackage{color}
\newcommand{\beq}{\begin{equation}}
\newcommand{\eeq}{\end{equation}}
\newcommand{\beqa}{\begin{eqnarray}}
\newcommand{\eeqa}{\end{eqnarray}}

\newcommand{\qvec}{{\bf q}}

\newcommand{\rvec}{{\bf r}}

\begin{document}
\title{Interplay between  density and superconducting quantum critical fluctuations}
\author{S. Caprara}
\affiliation{Dipartimento di Fisica, Universit\`a di Roma Sapienza, 
Piazzale Aldo Moro 5, I-00185 Roma, Italy}
\affiliation{Istituto dei Sistemi Complessi CNR and CNISM Unit\`a di Roma Sapienza}
\author{N. Bergeal}
\affiliation{LPEM UMR8213/CNRS - ESPCI ParisTech - UPMC PSL Research University, 10 rue Vauquelin - 75005 Paris, France}
\author{J. Lesueur}
\affiliation{LPEM UMR8213/CNRS - ESPCI ParisTech - UPMC PSL Research University, 10 rue Vauquelin - 75005 Paris, France}
\author{M. Grilli}
\affiliation{Dipartimento di Fisica, Universit\`a di Roma Sapienza, 
Piazzale Aldo Moro 5, I-00185 Roma, Italy}
\affiliation{Istituto dei Sistemi Complessi CNR and CNISM Unit\`a di Roma Sapienza}

\begin{abstract}
{We consider the case of a density-driven metal-superconductor transition in the proximity of an electronic phase separation.
In particular we investigate the interplay between superconducting fluctuations and density fluctuations, which become quantum critical 
when the electronic phase separation vanishes at zero temperature into a quantum critical point. In this situation
the critical dynamical density fluctuations strongly affect the dynamics of the Cooper pair fluctuations, which acquire a 
more singular character with a $z=3$ dynamical critical index. This gives rise to a scenario that possibly rules the disappearance
of superconductivity when the electron density is reduced by elecrostatic gating at the LaAlO$_3$/SrTiO$_3$ interface.}
\end{abstract}
\date{\today}
\maketitle
\section{Introduction}\label{intro}
There are many electronic systems where some kind of order is established upon changing the electron density. Traditional 
examples are given by disordered doped semiconductors or MOSFETs, where a metal-insulator transition is found upon reducing 
the electron density\cite{kravchenko} and by heavy fermion systems, where competing phases (like, e.g., antiferromagnetism) 
are often spoiled by doping, giving rise to metallic states often accompanied by superconductivity\cite{stewart}. High 
temperature superconducting cuprates and pnictides are other noticeable examples of systems where several electronic phases are realized
by changing doping. More recently it was also found that the two-dimensional electron gas (2DEG) formed at the interface of 
transition-metal oxides (like, e.g., LaXO$_3$/SrTiO$_3$, with X=Al,Ti, LXO/STO) become superconducting when the electron density 
is varied by electrostatic gating\cite{reyren,espci1}. A zero-temperature Quantum Critical Point (SC-QCP) separates the superconducting phase from the non-superconducting one\cite{caviglia2008}. It may happen that the electronic liquid is not homogeneously 
distributed in the system, as in the presence of an electronic phase separation (PS) for instance. In this case, 
if the electronic inhomogeneities are static and sizable, i.e., large enough to sustain the ordered phase (e.g., 
superconductivity) one may expect that an inhomogeneous ordered state can be realized.  For instance, one can envisage a state where superconducting "puddles" are embedded in a non-superconducting metallic phase of lower electronic density, and link together to establish an overall coherent state, which is destroyed when the electron density is reduced. \cite{SOK,espci_natmat}.

The phase separation regime itself may be tuned by an external parameter, such as an electrostatic gate voltage, 
which controls the average electronic density, and disappear into a zero-temperature critical point (PS-QCP).
 Around this point the density fluctuations have 
a quantum nature and keep a dynamical  character, giving rise to slow critical density fluctuations at very long wavelengths. Of 
course, this physical situation can only occur (at least approximately on large, even though not infinite, scales) if the 
Coulombic repulsion between electron density fluctuations is substantially screened. In this case the critical electron density 
fluctuations in the clean limit display a typical behavior characterized by their propagator\cite{CDG,maccarone}
\begin{eqnarray}
D(\qvec,\omega)&=&\langle \delta n(\qvec,\omega) \delta n(-\qvec,-\omega)\rangle\nonumber\\
&=&-\frac{1}{c_nq^2-i|\omega|/q+m_n^0}.
\label{npropagator}
\end{eqnarray}
where $c_n$ is a fermionic scale, $q\equiv |\qvec|$, and $m_n^0=c_n\xi_n^{-2}$ ($\xi_n$ being the density correlation length) 
rules the distance from the PS-QCP, where 
it vanishes. To be more specific, in LXO/STO this distance can be tuned by the gating potential $V_g$, $m_n^0 \propto (V_g^c-V_g)$,
where $V_g^c$ is the critical gating at which the QCP occurs.
Near this QCP the Landau damping induces a term $i|\omega|/q$. The largest contribution of these
fluctuations (i.e. the maximum of their spectral density) occurs when the damping term is of the same order of the $q^2$ term. 
This leads to an overdamped dynamics with $\omega\sim q^3$ and therefore a dynamical critical index $z=3$ characterizes the
critical density fluctuations.

Here, we address one single question: what happens if the QCP for phase separation occurs near the QCP of the 
density-driven ordering transition? For the sake of concreteness, we will explicitly consider the case of density-driven
superconductor-to-metal transition, but several of our arguments will apply to other transitions as well.
The rather generic finding is that the most critical (i.e., singular) fluctuations imprint the dynamics of the less critical 
fluctuations. In the specific case of PS-QCP and SC-QCP proximity we will see that the $z=2$ superconducting (Cooper) 
fluctuations inherit the $z=3$ dynamics of the underlying density fluctuations.

In order to provide a first intuitive glimpse of the situation, we first consider a toy model based 
on superconductivity instantaneously taking place as soon as a (dynamical but slow) density fluctuation increases locally the 
electron density above a critical value (Sect. II). In Sect. III we will provide general arguments to support the idea that the 
two QCP (the one closing the PS dome and the one related to density-driven superconductivity) tend to attract and coalesce. Sect. 
IV contains our concluding remarks.

\section{The fluctuating puddle model}
We start from a simple ``toy model'' model, where the critical character of the density-driven ordering is neglected, in order 
to make the above ``imprinting''  mechanism more explicit. We assume that when a certain critical density $n_c$ is locally reached, 
a superconducting (SC) order parameter instantaneously arises. In particular, it may happen that a dynamical density 
fluctuation is created, where locally the density is sufficiently high to sustain superconductivity. In this case, neglecting 
the transient behavior, one faces a superconducting puddle, which, however, only survives as long as the density fluctuation 
survives. It is then natural to infer that the dynamics of the density fluctuation imprints (actually it determines) the dynamics 
of the superconducting puddles and leads to an apparent $z=3$ behavior for SC dynamics. 

\begin{figure}
\includegraphics[angle=0,scale=0.3]{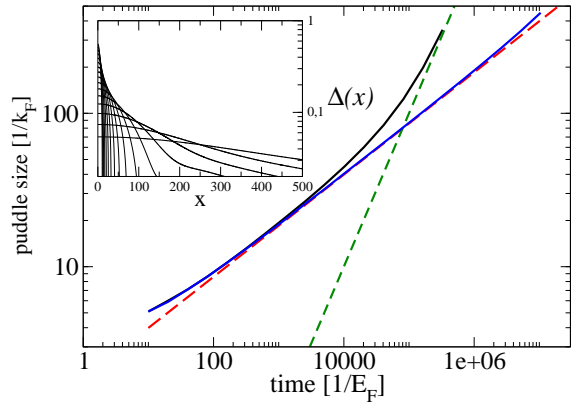}
\caption{Time dependence of the puddle size in the case of superconducting puddles at any positive density ($m_{sc}^0=0$). 
The long dashed (red) straight line identifies the slope $1/3$ characteristic of the 
$z=3$ diffusive dynamics. The short dashed (green)  straight line identifies the slope $1$ characteristic of the $z=1$ diffusive 
dynamics. The thick black curve is for $\xi_n^{-2}=10^{-3}$, while the thin blue curve is for $\xi_n^{-2}=10^{-5}$, in inverse 
of unit lengths 
square. Inset: spacial profile of the order parameter at various times, from $t=10$ to $t=10 \times 2^{14}$ (in $E_F^{-1}$ units). 
The initial size of the fluctuating puddle is of 3 unit lengths (in $k_F^{-1}$ units).}
\label{Deltaz3}
\end{figure}

Owing to the propagator in Eq. (\ref{npropagator}), the density fluctuations in momentum space [i.e., the Fourier transform of 
$\delta n(\rvec,t)\equiv n(\rvec,t)-\bar n$, $\bar n$ being the average density] satisfy the differential equation
\begin{equation}
\frac{1}{q}\partial_t \delta n(\qvec,t)=\left( \xi_n^{-2}+q^2 \right)\delta n(\qvec,t),
\label{nequation}
\end{equation}
where $\xi_n^{-2}=m_n^0/c_n$, and we take the momenta in units of Fermi wavevectors and time in units of inverse Fermi 
energy. Eq. (\ref{nequation}) can be solved giving, for $t>0$,
\begin{equation}
\delta n(\qvec,t)=\delta n(\qvec,t=0) e^{-q(\xi_n^{-2}+q^2)t}.
\label{ndit}
\end{equation}
The corresponding real-space density can be coupled to the SC order parameter  $\Delta (r)|$  by a term of the form 
\beq
H_{n\Delta}=-g\int   \delta n(r)|\Delta (r)|^2 \,d^dr,
\eeq
while the SC order parameter itself is ruled by a Hamiltonian
\[
H_{\mathrm{SC}}=\int \left[m_{sc}^0|\Delta (r)|^2+\frac{b}{2}|\Delta (r)|^4\right] \, d^dr,
\]
where $m_{sc}^0$ rules the distance from the SC critical point. Here
space and time dynamics of the SC order parameter are purposely neglected, within our toy model, in the absence
of coupling to the density fluctuations.
If the SC-QCP and the PS-QCP coincide at the same critical density $n_c$, and the SC phase is located at higher density,
a positive fluctuation of $\delta n=n-n_c>0$ instantaneously induces a static SC order parameter. On the other hand, if the 
PS-QCP occurs at slightly lower density $n_c^{ps}<n_c^{sc}$, one needs a sufficiently large density fluctuation 
$\delta n>n_c^{sc}-n_c^{ps}\equiv m_{sc}^0/g$ in order to create the SC order in the region of the fluctuation. Thus, inside the region 
with (sufficiently) positive $\delta n(\rvec,t)>m_{sc}^0/g$, one has
\begin{equation}
\left[\Delta(\rvec,t)\right]^2=\frac{g\delta n(\rvec,t)-m_{sc}^0}{b}\,\theta\left( g\delta n(\rvec,t) - m_{sc}^0 \right),
\label{delta}
\end{equation}
where $\theta(x)$ is the Heaviside function.
To test the occurrence of a $z=3$ dynamics for the SC order parameter, for the sake of definiteness, we consider the case
when superconductivity takes place in the puddle
as soon as $\delta n>0$ (i.e. $m_{sc}^0=0$). We numerically generated at $t=0$ in one dimension
a density fluctuation with a Gaussian space profile and, after Fourier transforming, we let it evolve in time according to 
Eq. (\ref{ndit}). Transforming back to real space, we insert the new density profile in Eq.(\ref{delta}) and we  
identify the time evolution of the frontier of the SC puddle, obtaining the 
curves of Fig. \ref{Deltaz3}.
One can see that, after a short time transient, a $z=3$ diffusive dynamics  (the long-dashed line is
a $y=1.85\,t^{1/3}$ curve for comparison) occurs with the front of the SC puddle increasing like
$r_\Delta \sim t^{1/3}$ (black and blue curves corresponding to $\xi_n^{-2}=10^{-3}$ and $\xi_n^{-2}=10^{-5}$ respectively).  
The inset reports the profile of the SC order parameters $\Delta(x)$ at various times in the case $\xi_n^{-2}=10^{-3}$. In this case
one can also see that, due to the finite extension of the critical density fluctuation ($\xi_n=10^{3/2}$ unit lengths), a long time crossover to a $z=1$ 
propagating behavior occurs for $t\gtrsim 10^5$ (short-dashed straight line).

A word is now in order to comment on the space and time units for systems like LXO/STO. The small densities involved 
($n\approx 0.01-0.08$ el/cell) are such that the Fermi energies are small (a few tens of meV) and the inverse Fermi wavevector is 
roughly of order of 5-20 lattice spacings. Thus the extension of the puddle before the $z=1$ crossover is reached can be estimated 
to be 300-3000 lattice units for $\xi_n^{-2}=10^{-3}$ or $\xi_n^{-2}=10^{-5}$, respectively. For instance, in LTO/STO 300 lattice units correspond to 
about 120 nm. 

\section{Perturbative hints}
\label{theory}
\begin{figure} 
\includegraphics[angle=0,scale=0.2]{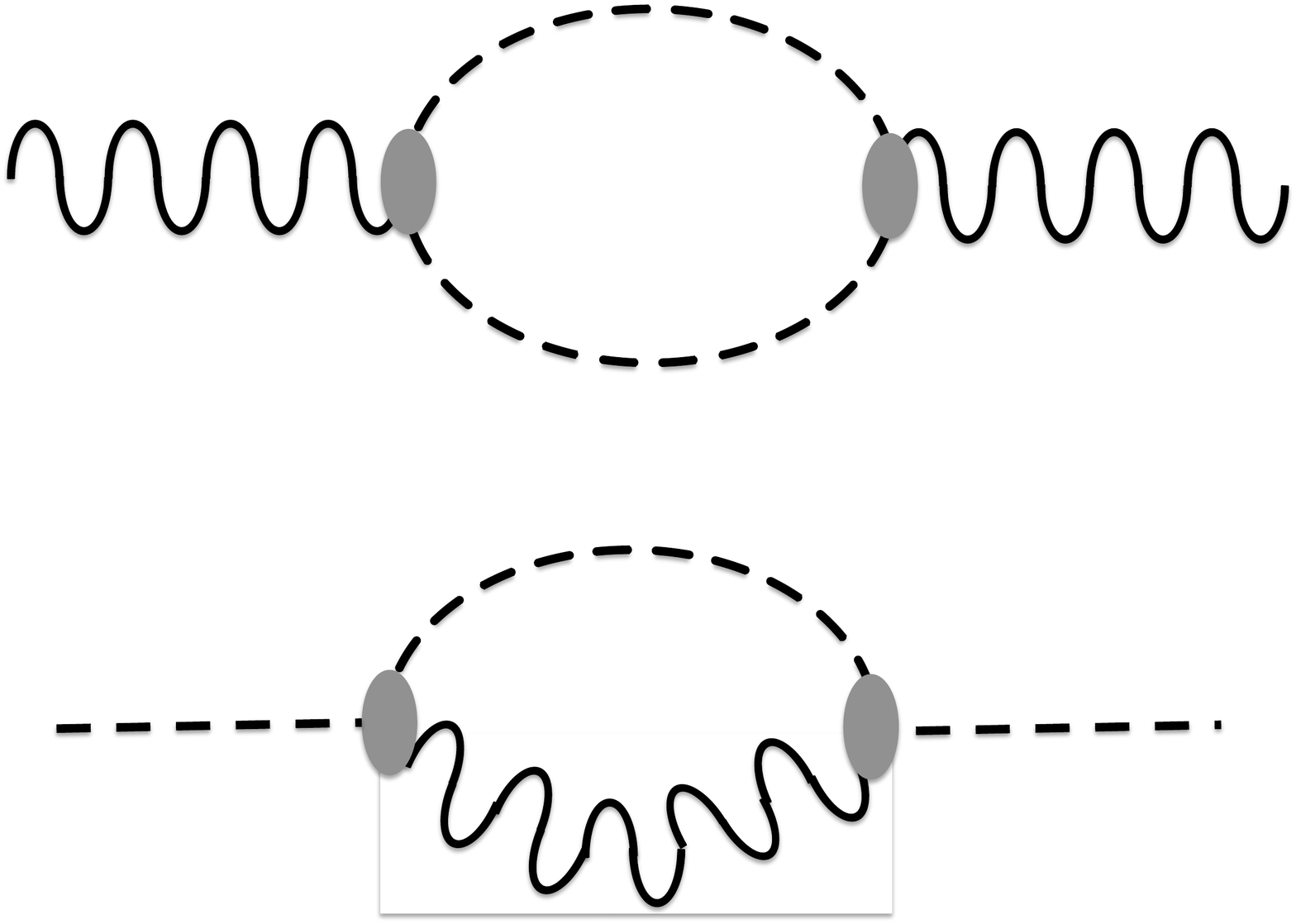}
\caption{(Above) Self-energy correction of the density-density propagator (wavy line) due to the Cooper fluctuations
(dashed lines), (below) Self-energy corrections of the Cooper propagator due to density fluctuations.}
\label{feynman}
\end{figure}

\begin{figure} 
\includegraphics[angle=0,scale=0.35]{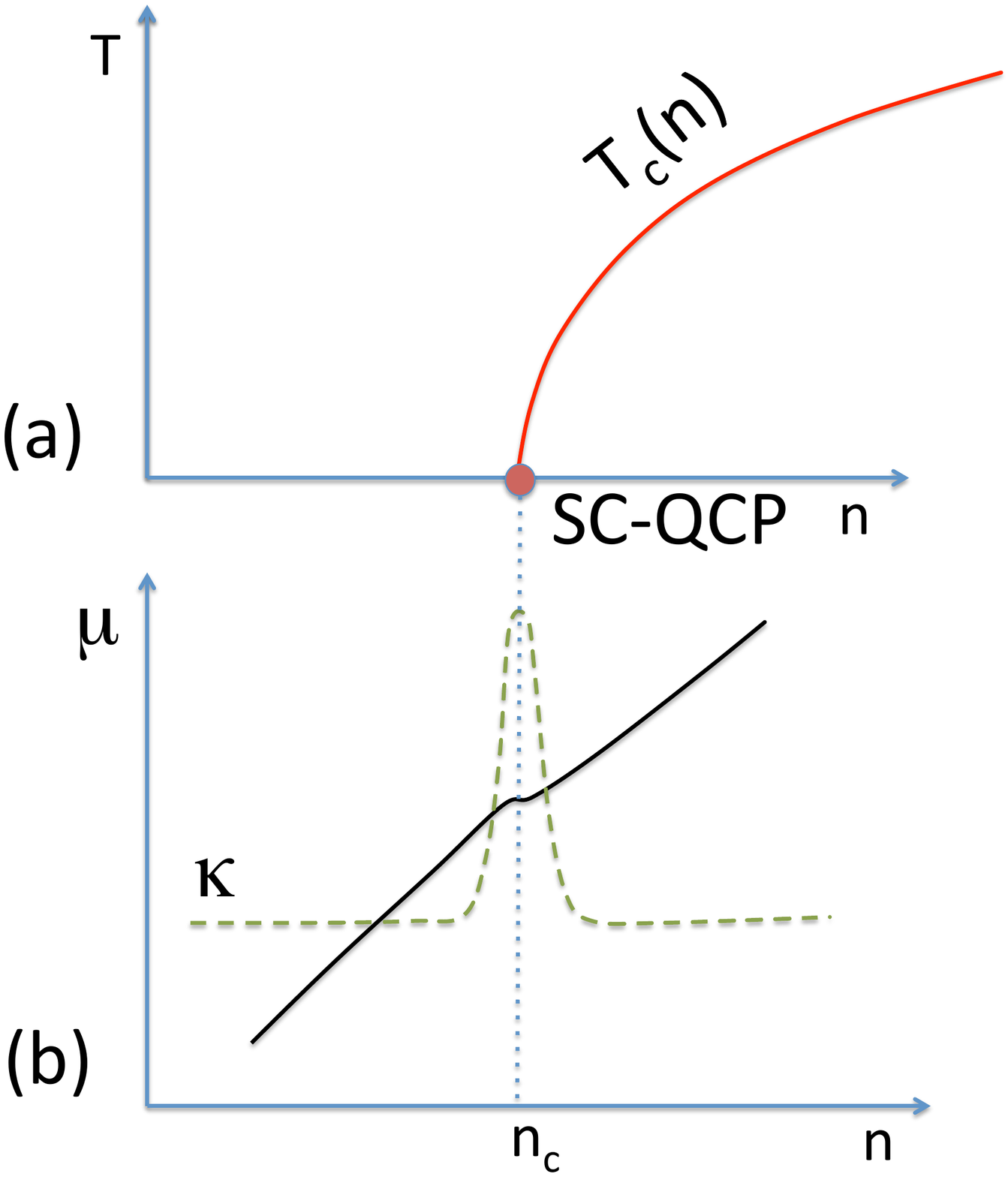}
\caption{(a) Schematic phase diagram of the density-driven superconductivity with a SC-QCP at a critical density $n_c$. 
(b) Chemical potential vs. density displaying an increase of the compressibility $\kappa$ (dashed line) in correspondence 
to the SC-QCP}
\label{compress}
\end{figure}

In the above model one finds that a static SC order parameter acquires a critical dynamics ``riding'' on a $z=3$ density fluctuations.
We now analyze from a simple perturbative point of view the interplay between the PS and the SC quantum criticality
recovering the critical character of the superconducting fluctuations, which was neglected in the above ``toy model''.
We consider the case when the superconducting order parameter fluctuates both in modulus and phase (like 
in the original Ginzburg-Landau case) and couples to nearly critical density fluctuations of an otherwise homogeneous metal.
This scenario involves the interplay between standard Cooper fluctuations and critical density fluctuations. An obvious 
prerequisite for this interplay is that the PS-QCP and the SC-QCP are in close proximity. This naturally raises the question 
whether such proximity requires a fine tuning of the model parameters or it naturally occurs due to some intrinsic ``attraction'' 
of the two critical point. The first result of our perturbative analysis is that the second possibility is quite likely. First of 
all, it is well known that SC pairing is favored (and it may even be induced) by critical density fluctuations. It has been shown 
in the framework of strongly correlated electrons that near a PS-QCP superconducting pairing is induced due to the mediation of 
the nearly critical density fluctuations\cite{GRCDK,perali}. On the other hand, it has also been proposed\cite{noiJSMM} that
a density-driven superconducting transition may favor phase separation. Here we perturbatively show that the density fluctuations are 
made softer (and the compressibility correspondingly increases) when the SC-QCP is approached. We calculate the diagram (a) in 
Fig. \ref{feynman}, representing the first perturbative corrections to $m_{n}^0$ in the $D$ propagator of Eq.(\ref{npropagator}) 
(which corresponds to the bare inverse compressibility). The dashed lines represent the usual Cooper fluctuation 
propagator\cite{varlamov}
\begin{eqnarray}
L(\qvec,\omega)&=&\langle \Delta(\qvec,\omega) \Delta(-\qvec,-\omega)\rangle\nonumber\\
&=&\frac{1}{c_{sc}^0q^2-i|\omega|+m_{sc}^0},
\label{cooper}
\end{eqnarray}
which clearly displays a $z=2$ ($\omega \sim q^2$) overdamped dynamics.
We find that, starting from separated QCPs, the finite bare inverse compressibility $m_n^0$  tends to be reduced by the coupling 
to the superconducting critical fluctuations
$$
m_n=m_n^0-g^2\log\frac{\Lambda}{m_{sc}^0}
$$
yielding a smaller dressed inverse compressibility $m_n$.
Here $m_n$ and $m_{sc}^0$  also measure the distances from the PS-QCP and SC-QCP respectively, while $\Lambda$ is a non-universal
high-energy cutoff. 
Therefore, no matter how weak is the coupling between superconducting and density fluctuations, the inverse compressibility 
$\kappa^{-1}=m_n$ tends to decrease when approaching the SC-QCP (where $m_{sc}^0\to 0$). A pictorial view of this scenario is
reported in Fig. \ref{compress}.
Although a full renormalization group calculation is in order to draw definite conclusions, the above perturbative finding
indicates that the proximity of the PS-QCP and the SC-QCP is not a fortuitous (and unlikely) coincidence.

Based on the above indication, we now investigate the interplay of the SC and density fluctuations assuming that the 
the SC and PS criticality are sufficiently close to strongly influence each other and we impose a formal coincidence of the 
two QCPs.  Thus we carry out perturbative calculations in two dimensions starting from the standard Cooper fluctuation propagator 
in Eq. (\ref{cooper}), with $z=2$, and the density fluctuation propagator of Eq. (\ref{npropagator}), with $z=3$. 
To impose the QCP coincidence maximizing the effect of the critical interplay we set to zero 
both $m_n^0$ and $m_{sc}^0$ in the calculation of the diagram in Fig. \ref{feynman} (b). We find  that the SC fluctuations are 
strongly renormalized and at first order their propagator acquires the form 
$$
\tilde{L}(\qvec,\omega)\approx
\frac{1}{c_{sc} q-i\omega^2/q^5}.
$$
Again the largest contribution of these fluctuations (i.e. the maximum of their spectral density) occurs when 
the damping term $i\omega^2/q^5$
is of the same order of the $q$ term leading to an overdamped dynamics with $\omega^2 \sim q^6$. And again this corresponds to a 
dynamical critical index $z = 3$. Notice that the constant terms in the denominator, which would drive the system away from the
critical point has been self-consistently set to zero to keep the criticality
condition fulfilled. Once more we stress that a full renormalization group treatment is required to firmly establish the behavior 
near the
two QCP, but our perturbative calculations are clearly indicative of the following scenario. The two QCP tend to coalesce 
and would generically give rise to a first-order transition with a PS between a low-density metallic state and a high-density 
SC state. However, as it usually occurs, a tuning of some parameter (like gating or magnetic field) may bring the system near a 
quantum tricritical point, where the SC critical temperature vanishes and, at the same time, the compressibility diverges, 
$\kappa=m_n^{-1} \to \infty$ (i.e., the two QCPs merge). At this point the SC fluctuations acquire a $z=3$ dynamics, 
while the effective dimensionality $d+z=5$ is larger than the upper critical dimension $d_{uc}$ which is three for a tricritical point.
In this case, for a two-dimensional system, one would expect mean-field critical indexes.
\begin{figure}
\includegraphics[angle=0,scale=0.3]{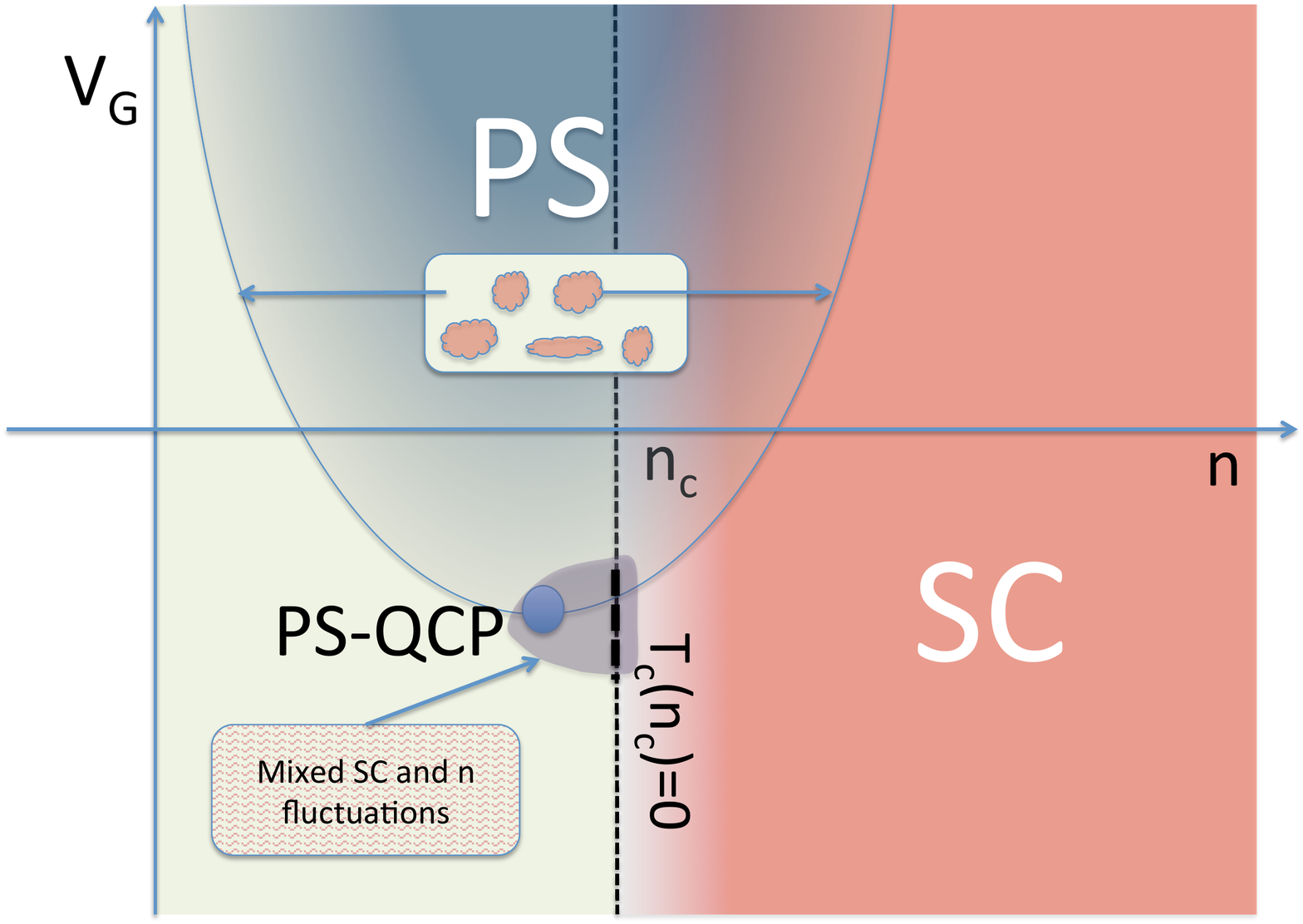}
%
\caption{Density $n$ vs. gating $V_g$ phase diagram. The SC critical temperature $T_c(n)$ vanishes along the line of QCP
represented by the black dotted line. In the PS region static puddles of denser superconducting phase are embedded in 
a less dense normal metallic phase.
The darker shaded area represents the region where PS-QCP interplay with the nearby SC-QCPs
giving rise to mixed dynamical quantum SC and density fluctuations.}
%
\label{ph-diag-B0}
\end{figure}

\section{possible realization: LXO/STO interface}
All the above analysis is based on the presence of long-ranged nearly critical density fluctuations. Obviously such fluctuations 
are only possible if the long-range Coulombic (LRC) electron-electron interaction is strongly screened, otherwise these fluctuations 
would become high-energy barely damped plasmonic excitations. In this case the dynamical critical index $z=1$.\cite{fisher,herbut}
 It has also been shown\cite{senthil} that a moderate screening, like the one due to a metallic gate close to a Josephson-Junction array,
may still lead to $z=1$ and a rather large value of the correlation length critical index $\nu>2$.\cite{nota_senthil} 
 The question therefore becomes: to what extent our strong-screening 
assumption is valid leading to the strong Landau-damping effects responsible for the large, $z=3$, value of the dynamical critical index? 
The answer naturally depends on the physical system at hand. If the SC transition occurs in a very low-density 
system, LRC interaction likely plays a relevant role and the scenario of the previous sections would hardly  apply.
On the other hand, when the SC transitions takes place in a metallic state with high mobility and relatively high electron densities, all interactions
become short ranged and the superconducting fluctuations may be strongly damped.  The text-book case of the metal-to-superconductor 
transition driven by Cooper fluctuations naturally realizes this case with $z=2$ [cf. Eq. (\ref{cooper})]. The same critical index
characterizes the superconducting QCP of Ref. \onlinecite{coleman}. 
LXO/STO interfaces, where $n\approx 10^{13}-10^{14}$ electrons/cm$^2$ likely belong to this class systems
where  the screening plays a relevant role thereby leaving the possibility of
large critical density fluctuations open. Moreover, one should also remember that SrTiO$_3$, where the interface charges reside,
has a huge dielectric constant ranging from 300 to 25000 strongly enhancing the screening.

Disorder is another relevant issue to be discussed. We consider systems where the microscopic disorder is weak and rather 
homogeneously distributed. In this case the metallic character of the systems  is poorly affected by the impurity scattering 
at most resulting in weak localization of no relevance to our purposes (like in LTO/STO). On the other hand, the presence of a PS 
naturally entails the occurrence of a strongly inhomogeneous state emerging from this otherwise weakly disordered metal. This 
inhomogeneity would poorly affect transport properties because it mostly enhances forward scattering leaving the mobility 
high. However such static inhomogeneity would greatly affect the critical properties raising the question of the Harris criterion to 
be satisfied in order to have a true phase transition.
The metallic state near the PS-QCP is not disordered because the density fluctuations are slow (that is critical),  but fully
dynamical excitations of the electron gas and they are not a static (quenched) source of disorder. In this case the Harris 
criterion simply does not apply (at least, in its standard form). 
In particular, when the metal-SC transition is driven by the gating potential $V_g$ a $z\nu=3/2$ scaling is observed.\cite{espci_Vg}
In this case one can argue that gating brings the system close to a PS-QCP,
where no static puddles are present (thus  Harris criterion is not applicable) and the density critically fluctuates in a clean metallic state.
In this scheme the experimental observation of $z\nu=3/2$ could arise according to the scenario of 
Sect. \ref{theory}  from a $z=3$ dynamical critical 
behavior of the superconducting fluctuations together with a mean-field like exponent $\nu=1/2$. 
 We find it appealing to consider and explore the possibility of this alternative scenario, in which 
the superconducting fluctuations are coupled to (and imprinted by) nearly critical density fluctuations.

We finally remark that our scenario may also apply to those cases where superconducting fluctuations couple to longitudinal nematic
modes (also characterized by $z=3$\cite{garst}) possibly explaining the occurrence of a superconducting QCP with $z\nu=3/2$
reported in a gate doped La$_{2-x}$Sr$_x$CuO$_4$ sample.\cite{bollinger}

The Authors gratefully thank C. di Castro, C. Castellani, L. Benfatto for stimulating discussions. This work has been supported by the Region Ile-de-France in the framework of CNano IdF, OXYMORE
and Sesame programs, by DGA and CNRS through a PICS program (S2S) and ANR JCJC (Nano-SO2DEG). 
S.C. and M.G. acknowledge financial support form the University of Rome Sapienza with the Project n. C26A125JMB.


\begin{thebibliography}{99}
\bibitem{nota_senthil} In Ref. \onlinecite{senthil} the case of $1<z<2$ was also consider, but the values of $\nu$
were not reported in this case.
\end{thebibliography}
\end{document}